\documentclass{ws-procs9x6}

\begin{document}
\title{	  Description of the dynamics of a random chain with rigid
  constraints in the path integral framework
}

\author{Franco Ferrari$^*$ and Jaros{\l}aw Paturej$^\dagger$}

\address{Institute of Physics, University of Szczecin,\\
Wielkopolska 15, 70451 Szczecin, Poland\\
$^*$E-mail: ferrari@fermi.fiz.univ.szczecin.pl\\
$^\dagger$ E-mail: jpaturej@univ.szczecin.pl}

\author{Thomas A. Vilgis$^\ddagger$}
\address{Max Planck Institute for Polymer Research,\\
 10  Ackermannweg, 55128 Mainz, Germany\\
E-mail: vilgis@mpip-mainz.mpg.de }

\begin{abstract}
In this work we discuss the dynamics of a three dimensional
chain.
It turns out that
the generalized nonlinear sigma model presented in
Ref.~\refcite{FePaVi} may be easily generalized to three dimensions. The
formula of the probability distribution of two topologically 
entangled chain is provided. The interesting case of a chain which can
form only discrete angles with respect to the $z-$axis is also
presented. 
\end{abstract}


\bodymatter
\section{Introduction}
In this brief report the dynamics of a chain fluctuating in some
medium at fixed temperature $T$ is discussed.
The three dimensional case is particularly interesting, because it
allows to study the topological entanglement of two or more
chains. 
The problem of the topological entanglement of two chains has been
investigated for a long time in the statistical mechanics of
polymers, see for instance Ref.~\refcite{vilreview} and references
therein. If 
the topological constraints which limit the fluctuations of
the chains are described by using the Gauss linking number, the
probability distribution of the system turns out to be equivalent to
the partition function of a zero-component Landau--Ginzburg model
interacting with a pair of Chern--Simons fields. 
The analogous problem in polymer dynamics has not yet been solved.
Here we show that the probability function of the system of two
chains in the presence of topological constraints may be simplified
thanks to the introduction of a Chern--Simons field theory also in the
 case of dynamics. 
Finally, we provide a  formula for the probability function
of a chain which can
form only discrete angles with respect to the $z-$axis.
\section{A model of two topologically entangled chains
}\label{aba:sec1}  
We would like to treat the dynamics of a chain fluctuating in
some medium at constant temperature $T$. In Ref.~\refcite{FePaVi} (see also
Ref.~\refcite{FePaViproc}) the case of a two dimensional
chain has been discussed. The approach presented in those references can
however be extended to any dimension. Let us consider the
 probability function $\Psi$ which measures the probability that a
$D-$dimensional continuous chain starting from a given spatial
 configuration $\mathbf 
 R_i(s)$ arrives after a time $t_f-t_i$ to a final configuration
 $\mathbf R_f(s)$. 
The chain is regarded as the continuous limit of a discrete chain
consisting of particles connected together with segments of fixed
length.
In the continuous limit, the constraints
arising due to the presence of the segments take the form:
$
\left|
\frac{\partial \mathbf R(t,s)}{\partial s} 
\right|^2=1
$.
Then, an expression of $\Psi$ in terms of path
 integrals may be written as follows:
\begin{eqnarray}
\Psi%[\mathbf R_f(s),\mathbf R_i(s)]
&=&\int_{\mathbf R(t_f,s)=
\mathbf R_f(s)\atop
\mathbf R(t_i,s)=
\mathbf R_i(s)
}
{\cal D}\mathbf R{\cal D}\lambda\exp\left\{
-c{\textstyle
\int_{t_i}^{t_f}dt\int_0^Lds\dot{\mathbf
  R}^2
}
\right\}
\nonumber\\
&\times&\exp{\textstyle\left\{i\int_{t_i}^{t_f}dt\int_0^Lds\lambda\left(
{\textstyle\left|
\frac{\partial \mathbf R}{\partial s} 
\right|^2-1}
\right)\right\}}
\label{contparttd}\end{eqnarray}
where the fields $\mathbf R(t,s)$ represent $D-$dimensional vectors.
Moreover $c=\frac{M}{4kT\tau L}$,
$M$ is the total mass of the chain, $L$ is its
length and $k$ denotes the Boltzmann constant. Finally, the
 relaxation time $\tau$
characterizes the rate of the decay of the drift velocity of the
particles composing the chain.
In Eq.~(\ref{contparttd}) 
the Lagrange multiplier $\lambda$ has been introduced in order to
impose the constraints using the Fourier representation of
 the $\delta-$function $\delta(\left|
\frac{\partial \mathbf R}{\partial s} 
\right|^2-1)$.
The model described by
Eq.~(\ref{contparttd}) will be called here the
generalized nonlinear sigma model (GNLSM) due to its close resemblance
to a 
two-dimensional nonlinear sigma model.
Let us note that the holonomic constraint $\mathbf R^2=1$ of the
nonlinear sigma model has been replaced here by a nonholonomic
constraint. 

In the following, we will restrict ourselves to the physically
relevant case $D=3$.
The three dimensional case is particularly interesting because it
allows the introduction of topological relations.
To this purpose, let us imagine two closed chains $C_1$ and $C_2$ of
lengths $L_1$ and $L_2$ respectively. The trajectories of the two
chains are described by the two vectors $\mathbf R_1(t,s_1)$ and
$\mathbf R_2(t,s_2)$ where $0\le s_1\le L_1$ and $0\le s_2\le L_2$.
The simplest way to impose topological constraints on two closed
trajectories is to use the Gauss linking number $\chi$:
$\chi(t,C_1,C_2)=\frac{1}{4\pi}\oint_{C_1}d\mathbf R_1\cdot\oint_{C_2}
d\mathbf R_2\times\frac{(\mathbf R_1-\mathbf R_2)}{|\mathbf
  R_1-\mathbf R_2|^3}$.
If the trajectories of the chains were impenetrable, then $\chi$
would  not depend on time, since it is not possible to change the
topological configuration of a system of knots if their trajectories
are not allowed to cross themselves.
However, since we are not going to introduce interactions between the
two chains, we just require that, during the time $t_f-t_i$, the
average value of the Gauss linking number is an arbitrary constant
$m$, i.~e.
$m=\frac 1{t_f-t_i}\int_{t_i}^{t_f}\chi(t,C_1,C_2)dt$.
It is possible to show that the probability function of two chains
whose trajectories  satisfy the above topological
constraint is given by:
\begin{equation}
\Psi(C_1,C_2)=\int {\cal D}(fields)e^{-(S_1+S_2)}e^{iS_{CS}+\frac{i\mu} 2
\int_{-\infty}^{+\infty}d\xi\int d^3x\mathbf J^{i}\mathbf A_i
}
\end{equation}
where ${\cal D}(fields)=\prod_{i=1}^2{\cal D}\mathbf R_i{\cal D}\lambda_i
{\cal D}\mathbf A_i$,
\begin{equation}
S_i={\textstyle \int_{t_i}^{t_f}dt\int_0^Lds_i\left[
c\dot{\mathbf R}^2_i+i\lambda_i\left(
\left|
\frac{\partial \mathbf R_i}{\partial s}
\right|^2-1
\right)
\right]}\quad i=1,2
\end{equation}
\begin{equation}
S_{CS}=\frac 1{t_f-t_i}\int_{-\infty}^{+\infty} d\xi\int d^3 x \mathbf
A_1(\xi,\mathbf 
x)\cdot(\mathbf \nabla_{\mathbf x}\times\mathbf A_2(\xi,\mathbf x))
\end{equation}
\begin{equation}
\mathbf J^i(\xi,\mathbf x)=\int_{t_i}^{t_f} dt\int_0^{L_i}
ds_i\delta(\xi -t) \frac{ \partial \mathbf R_i(t,s_i)}{\partial s_i}
\delta^{(3)}(\mathbf x-\mathbf R_i(t,s_i))\quad i=1,2
\end{equation}
\section{Chain with constant angles of bending}
To conclude, we would like to mention the interesting case in which
the chain is forced to form with the $z-$axis only the two fixed
angles $\alpha$ and $\pi -\alpha$\cite{FePaVi1}. If there are no
interactions depending on the $z$ degree of freedom, it turns out that
this problem can be reduced to a two dimensional one. Since in this work 
 interactions are not considered, the probability
function of the chain may be written as follows:
\begin{equation}
\Psi^{3d}_{\alpha,\pi-\alpha}
=\int{\cal D}x{\cal D}y \exp\left\{-S_{\alpha,\pi-\alpha}
\right\}\delta(
(\partial_s x)^2+(\partial_s y)^2-\tan^2\alpha
)
\label{3ddirfiwn}
\end{equation}
where
$
S_{\alpha,\pi-\alpha}=c\sin^2\alpha\int_{t_i}^{t_f}
dt\int_0^Lds\left[  
\dot x^2+\dot y^2
\right]$.
%\label{djfjskdfldws}
%\end{equation}
\section{Conclusions}
In this work the dynamics of a $D-$dimensional chain has
been investigated. The probability function $\Psi$ of
this system is 
equivalent to the partition function of a generalized nonlinear sigma
model.
Next, the fluctuations of two topologically entangled chains have been
discussed. Analogously to what happens in the case of  statistical
mechanics\cite{fereview,topstamec}, the complexities
connected 
with the 
handling of the Gauss linking number may be partly eliminated with the
introduction of Chern-Simons fields, which decouple the interactions
of topological origin
between the chains \cite{Fenova}. Still, one has to perform a path
integration over the trajectories of each chain separately.
In statistical mechanics, this is equivalent to compute the path
integral of a particle immersed in a magnetic field. 
In dynamics, the particle is replaced by a 
two dimensional field $\mathbf R(t,s)$. To evaluate such path
integral is a
complicated task.
Finally, the problem of a three dimensional chain admitting only fixed
angles with respect to the $z-$axis is reduced to the problem of a two
dimensional chain, in a way which is similar to the reduction of the
statistical mechanics of a directed polymer to the random walk of a
two dimensional particle.
\section{Acknowledgments} 
This work has been financed in part
by the Polish Ministry of Science, scientific project N202 156
31/2933 and by the action
COST~P12 financed by the European Union. 
F. Ferrari and J. Paturej are also grateful to W. Janke, A. Pelster
and to the Max Planck Institute for the Physics of Complex Systems
for the nice hospitality in Dresden.

\end{document}